\documentclass[sn-mathphys-num]{sn-jnl}
\pdfoutput=1
\usepackage{graphicx}%
\usepackage{multirow}%
\usepackage{amsmath,amssymb,amsfonts}%
\usepackage{amsthm}%
\usepackage{mathrsfs}%
\usepackage[title]{appendix}%
\usepackage{xcolor}%
\usepackage{textcomp}%
\usepackage{manyfoot}%
\usepackage{booktabs}%
\usepackage{algorithm}%
\usepackage{algorithmicx}%
\usepackage{algpseudocode}%
\usepackage{listings}%
\usepackage{cancel}

\def\beq{\begin{equation}}
\def\eeq{\end{equation}}
\def\bea{\begin{eqnarray}}
\def\eea{\end{eqnarray}}

\def \pa {\partial}

\def \Ncal {\mathcal{N}}

\def \Ucal {\mathcal{U}}
\def \Nbcal {\bar{\Ncal}}

\title[$\delta\mathcal{N}$ formalism on the past light-cone]{$\delta\mathcal{N}$ formalism on the past light-cone}

\author[1]{\fnm{Giuseppe} \sur{Fanizza}}\email{fanizza@lum.it}
\author[2]{\fnm{Giovanni} \sur{Marozzi}}\email{giovanni.marozzi@unipi.it}
\author[2,3]{\fnm{Matheus} \sur{Medeiros}}\email{mrmsilva@fc.ul.pt}

\affil[1]{Dipartimento di Ingegneria, Universit\`a LUM,
S.S. 100 km 18 - 70010 Casamassima (BA), Italy}
\affil[2]{Dipartimento di Fisica, Universit\`a di Pisa, Largo B. Pontecorvo 3, 56127 Pisa, 
Italy,
and Istituto Nazionale di Fisica Nucleare, Sezione di Pisa, Italy}
\affil[3]{Instituto de Astrof\'isica e Ci\^encias do Espa\c{c}o, Faculdade de Ci\^encias da Universidade de Lisboa, Edificio C8, Campo Grande, P-1749-016, Lisbon, Portugal}

\abstract{We apply the gradient expansion approximation to the light-cone gauge, obtaining a separate universe picture at non-linear order in perturbation theory within this framework. Thereafter, we use it to generalize the $\delta\mathcal{N}$ formalism in terms of light-cone perturbations. As a consistency check, we demonstrate the conservation of the gauge invariant curvature perturbation on uniform density hypersurface $\zeta$ at the completely non-linear level. The approach studied provides a self-consistent framework to connect at non-linear level quantities from the primordial universe, such as $\zeta$, written in terms of the light-cone parameters, to late time observables.}

 
\begin{document}

\maketitle

\section{Introduction}

Advances in cosmological observations have provided us with high-precision methods to study the universe \cite{Abate:2012za,Amendola:2016saw,Aghamousa:2016zmz}. So far, linear cosmological perturbation theory has been the main tool to describe the early universe, particularly the primordial seeds that are believed to be produced by quantum mechanical fluctuations during inflation. These fluctuations grow during the quasi-exponential expansion epoch and freeze outside the horizon. Later on, they re-enter the horizon during a power-law expansion epoch, giving rise to the large-scale structure observed in the universe.

In order to link the gauge invariant quantity that characterizes such primordial fluctuations with the observations, we need to have a good understanding of their behaviour outside the horizon. An interesting example is given by the primordial curvature perturbation on the uniform density hypersurface $\zeta$, which is expected to be of order $10^{-5}$ at the last-scattering surface and has been shown to be conserved outside the horizon, both at the linear \cite{Wands:2000dp} and the non-linear level \cite{Lyth:2004gb}. The first-order treatment for the primordial fluctuations agrees with the observations of a nearly Gaussian, scale-invariant power spectrum. Although non-linearities are expected to be small, they are however unavoidable as a consequence of the non-linear evolution of the perturbations. Detection of the related non-Gaussianities can then provide important insights into early universe models, such as the inflationary ones \cite{Cai:2010uw,Komatsu:2000vy}.

The evolution of $\zeta$ is proportional to the non-adiabatic (if any) contribution in the energy-momentum tensor, as shown in \cite{Wands:2000dp, Lyth:2004gb}. In the linear regime, $\zeta$ has been successfully calculated using the $\delta\mathcal{N}$ formalism \cite{Sasaki:1995aw,Sasaki:1998ug,Salopek:1990jq,Starobinsky:1982ee,Starobinsky:1986fxa,Lyth:2005fi}, which has been extended to the exact non-linear level \cite{Lyth:2004gb} by applying the first-order gradient expansion directly in the equations of motion provided by the Arnowitt-Deser-Misner (ADM) formalism \cite{Arnowitt:1962hi}. The first-order gradient expansion, also known as the separate universe (SU) scheme, describes the universe as a set of FLRW geometries with independent equations of motion and is a good approximation in the regime of large comoving wavelengths compared to the horizon \cite{Sasaki:1995aw,Sasaki:1998ug,Wands:2000dp,Lyth:2004gb}.

The great advantage of the SU scheme is that the equations of motion within this approximation have the same form both for the background and perturbed universe with the exception of the momentum constraint, which vanishes in the background. As a consequence, one can obtain the non-linear field's evolution from the background one by imposing non-linear initial conditions \cite{Sugiyama:2012tj}.

In \cite{Vennin:2015hra}, this formalism has been generalized to include stochastic effects and derive, within the framework of the stochastic approach \cite{Starobinsky:1986fx} and its relation with QFT \cite{Finelli:2008zg, Finelli:2010sh}, non-perturbative correlation functions for single-field slow-roll inflation. Further extensions have also studied ultra-slow-roll inflation \cite{Prokopec:2019srf,Firouzjahi:2018vet,Ballesteros:2020sre,Pattison:2021oen}, allowing for the investigation of primordial black-hole production \cite{Prokopec:2019srf,Firouzjahi:2018vet,Ballesteros:2020sre,Pattison:2021oen,Biagetti:2018pjj,Ezquiaga:2018gbw}. The $\delta\mathcal{N}$ formalism has also been extended to the case when cosmic shear is included to describe the anisotropic expansion. In such a framework, the evolution of gravitational waves has been explored both for the case of a Bianchi I universe \cite{Talebian-Ashkezari:2016llx} and when couplings with external fields are present \cite{Tanaka:2021dww, Tanaka:2023gul}.

A formalism that connects the picture of the primordial universe presented so far to late-time observables would be greatly welcome, especially if such a connection can account for non-linearities. Since the Geodesic Light-Cone (GLC) gauge \cite{Gasperini:2011us} gives the chance of describing light-like observables in the late universe exactly, such as the redshift and the distance-redshift relation \cite{Gasperini:2011us,BenDayan:2012pp,BenDayan:2012ct,BenDayan:2012wi,Fanizza:2013doa,Marozzi:2014kua,Fanizza:2019pfp,Fanizza:2020xtv}, the galaxy number count \cite{DiDio:2014lka,DiDio:2015bua}, the non-linear corrections to the CMB spectra \cite{Marozzi:2016uob,Marozzi:2016qxl}, and also Ultra-Relativistic particles \cite{Fanizza:2015gdn}, this is a natural framework to pursue the aforementioned program.

Moreover, there has also been recent interest in the GLC gauge application to the study of backreaction effects from the primordial universe \cite{Fanizza:2020xtv,Fanizza:2023ixk,Frob:2021ore,Mitsou:2020czr}. Although these are very interesting prospects, one still must face the fact that the evolution of perturbations in the GLC gauge is quite involved already at linear order (see \cite{Mitsou:2020czr} for an analytical treatment). An alternative approach to this may be provided by numerical attempts, as done for instance in \cite{Tian:2021qgg} for the linearized evolved solution for the gravitational potential on the past light-cone. In this manuscript, we take a different route by providing simplified equations of motion using the SU approach on the past light-cone.

The connection between the primordial origin of inhomogeneities and their observations has to deal with the fact that the latter are done along our past light-cone, whereas the primordial universe is usually described using spatial hypersurfaces. During primordial epochs, these hypersurfaces are naturally described in terms of uniform field slices. In fact, in a single-field inflationary scenario, the inflaton is the only clock available. Therefore, the natural slicing which describes the dynamical space-time evolution is the one given by uniform inflaton hypersurfaces, which also fixes the time gauge mode. Another interesting fixing for the time coordinate is the one describing uniform density slices. This is an interesting fixing because it directly translates the density perturbations into curvature perturbations, providing the initial conditions for large-scale structure formation. These two gauge fixings are usually called, respectively, uniform field gauge (UFG) and uniform density gauge (UDG).

To make contact between these gauges and the GLC one, we recall that, although the GLC time coordinate is fixed to the time measured by a free-falling observer, a generalization of this gauge is provided in \cite{Mitsou:2020czr}, the so-called Light-Cone (LC) gauge. In this generalization, the time gauge choice is left unspecified, allowing us to fix the lapse function for describing the uniform field and uniform density slicing on the past light-cone. An alternative approach could be to start from the cosmological perturbation theory on the past light-cone developed in \cite{Fanizza:2020xtv,Fanizza:2023ixk}, and then perform the necessary gauge transformations. A description of the primordial universe in terms of the LC gauge could be a promising framework to connect non-linearly the late universe to the primordial one, given in terms of light-cone parameters, since cosmological observables can be described non-perturbatively within this gauge.

Here, we will take a first step in  this program. Moreover, we underline that the success of this program would probably require also numerical studies besides the analytical ones developed so far, regarding the late time evolution of the perturbations. In particular, along this manuscript, we will discuss the gradient expansion as done on the observer's past light-cone, which allows us to obtain a SU picture and the $\delta\mathcal{N}$ formalism in terms of light-cone perturbations. We will provide this both at the fully non-perturbative level, using the LC gauge \cite{Mitsou:2020czr}, and, as a consistency check, at the linear level using the light-cone perturbation theory \cite{Fanizza:2020xtv,Fanizza:2023ixk}.
By considering the LC gauge as a non-linear ADM decomposition (see \cite{Mitsou:2020czr} for more details), we will show that, unlike previous literature, where the shift vector was a first-order term in the gradient expansion \cite{Sugiyama:2012tj,Lyth:2004gb,Talebian-Ashkezari:2016llx}, in the LC gauge the shift vector has to be taken into account also for the background. This is an important difference, since in this case the shift vector corresponds to the direction of propagation of the photon, and it is used to take into account inhomogeneities along the photon propagation direction. However, we will neglect spatial derivatives of such shift vector since they correspond to light-cone distortion effects which are expected to be negligible on large scales.

After such implementations, we will show that the SU picture can be realized in the LC gauge (i.e., we will obtain evolution equations with the same form for both perturbed and background universe). Furthermore, within the gradient expansion approximation, we will verify at the fully non-linear level that the curvature perturbation on uniform density slices $\zeta$ is a conserved quantity (for adiabatic pressure) also when the light-like slicing of spacetime is used. This is a sanity check that confirms how the SU picture can be extended also to the case of the light-cone gauge.

In summary, we will present a novel approach to study the primordial universe within the past light-cone, by developing the $\delta\mathcal{N}$ formalism in the LC gauge. We will verify that the LC gauge allows for a non-linear description of the primordial universe in terms of light-cone parameters, and that the SU picture can be realized in this gauge by neglecting spatial derivatives of the shift vector.

The manuscript is organized as follows. In Sect.~\ref{SecConsXi} we obtain a generic SU description where we keep inhomogeneities along the geodesics in terms of an ADM metric. 
In Sect.~\ref{LCgauge} we present the set of light-cone gauges used here and, with a non-linear diffeomorphism, we show how the standard ADM formalism relates with the LC gauge. Moreover, we provide the LC gauge fixing condition in terms of the ADM variables up to non-linear order. Thereafter, we discuss the SU formalism on the past light-cone, which is also presented for the GLC gauge. Sect.~\ref{deltaN} is devoted to the computation of the non-linear number of e-folds and its relation to spacetime perturbations, which allow us to obtain the non-linear scale factor in the LC. We then give a proof for the super-horizon conservation of $\zeta$, at both linear and non-linear order in perturbation theory, and first order in the gradient expansion, for adiabatic fluids. Finally, we provide a generalization of the $\delta\mathcal{N}$ formalism in the LC gauge. In Sect.~\ref{Conclusions}, our main conclusions are summarized and discussed.
In Appendix \ref{secNvsGLC} we provide the linear $\delta\mathcal{N}$ formalism as a consistency check of the obtained results.


\section{Separate universe}
\label{SecConsXi}

Let us begin by introducing a systematic approximation scheme, which can be used when the wavelength of the perturbations is larger than the physical horizon. This approximation, widely known as the SU approach \cite{Sasaki:1995aw,Sasaki:1998ug,Wands:2000dp,Lyth:2004gb}, consists of employing the already mentioned gradient expansion perturbative scheme. This is based on the quantity $\epsilon\equiv k/(aH)$, rather than on the amplitude of the perturbations. This quantity compares the comoving wavenumber of a given mode $k/a$ with the expansion rate $H$. As an example, within this approximation scheme, terms with one spatial derivative will be first order\footnote{As an example in a flat space, under a Fourier transformation, spatial derivatives give rise to terms proportional to $k$. Here we are considering that for a quantity $Q$, $\frac{1}{a} \partial_{i}Q\ll\partial_{t}Q\approx HQ$
\cite{Salopek:1990jq}.} in $\epsilon$. The first order gradient expansion is known as the SU approach, since in this case the equations of motion for a local patch of the perturbed space-time have the same form of the FLRW background ones \cite{Sugiyama:2012tj}. Thus, in this view the universe can be described as a collection of FLRW geometries, each one locally described by a different scale factor.

This approximation can be particularly interesting to study the super-horizon evolution of the curvature perturbation $\zeta$ with a light-cone foliation of the spacetime. In fact, the SU approach is used
in \cite{Wands:2000dp,Lyth:2004gb} to show the conservation
of $\zeta$ on super-horizon scales, for adiabatic pressure, at linear
and non-linear order in perturbation theory.

By applying the gradient expansion to the non-linear ADM formalism \cite{Sugiyama:2012tj} one can provide
a SU scheme in the uniform curvature gauge (UCG). It has been shown that the shift vector has a decaying evolution, and therefore during the exponential expansion of the universe, it can be considered as a first order term in the gradient expansion. Also analyzing the consistency between the Hamiltonian and momentum constraints, it has been shown that, taking into account also the momentum constraints, the results differ only by a decaying solution \cite{Sugiyama:2012tj} (see also \cite{Garriga:2015tea} for a similar proof beyond the context of slow-roll inflation). Moreover, it is shown that the additional information in the momentum constraints should be $\mathcal{O}(\epsilon^{3})$ in the gradient expansion. Thereby, for super-horizon perturbations, the SU scheme is a good approximation.

In this manuscript, we will provide a SU picture for the LC \cite{Mitsou:2020czr} and GLC \cite{Gasperini:2011us} gauges. As we will see later, one difference with the previous works is that, when we consider the LC gauge as an ADM decomposition, the shift vector does not vanish, not even on the background (see, for instance \cite{Mitsou:2020czr}). In fact, in the LC gauge, the shift vector describes the direction of observation. On the other hand, we will neglect the divergence of the shift vector, which describes the divergence of the direction of observation in the language of $1+3$ formalism.


\subsection{ADM formalism}
\label{SADMU}

In this section we provide the SU set of equations for generic perturbations. Firstly, we introduce the ADM splitting and the $1+3$ evolution equations, then we obtain general conditions which allow a SU evolution of the perturbations. For a general formulation of the SU approach in the Hamiltonian formalism, we redirect the reader to \cite{Artigas:2021zdk}. Thereafter, we show how also the LC gauge satisfies these conditions. Starting
with the ADM metric
\begin{equation}
ds_{ADM}^{2}=-\mathcal{M}^{2}dt^{2}+f_{ij}\left(dx^{i}+N^{i}dt\right)\left(dx^{j}+N^{j}dt\right)\,,\label{eq:ADM1}
\end{equation}
one can prove that, with a suitable choice of the coordinates,
$N^{i}=\mathcal{O}\left(\epsilon\right)$ and therefore $\partial_{i}N^{i}=\mathcal{O}\left(\epsilon^{2}\right)$.
This condition was assumed in the references \cite{Lyth:2004gb,Wands:2000dp},
and it was proved in \cite{Sugiyama:2012tj} considering the UCG.

Let us now work with the ADM foliation of Eq.~\eqref{eq:ADM1}. Thanks to this description of non-linear general perturbations, made on top of a FLRW background, we will show how to recover a SU picture even if the shift vector does not vanish in the background. It rather combines with the time derivative to provide a derivative along the time-like motion.

The vector $n^\mu$ normal to the space-like hypersurfaces $t=const$ is given by
\begin{equation}
n^{\mu}=\frac{\partial^{\mu}t}{\left(-\partial_{\nu}t\partial^{\nu}t\right)^{\frac{1}{2}}}\,,\label{eq:def-n}
\end{equation} 
which satisfies
\begin{equation}
n_{\mu}=-\mathcal{M}\delta_{\mu}^{t}\,,\qquad n^{\mu}\partial_{\mu}=\frac{1}{\mathcal{M}}\left(\partial_{t}-N^{i}\partial_{i}\right)\,,
\end{equation}
with correspondent induced metric given by
\begin{equation}
f_{\mu\nu}=g_{\mu\nu}+n_{\mu}n_{\nu}\,.\label{eq:orthogonal-projector}
\end{equation}
This metric can be used to define the following induced quantities
\begin{equation}
E\equiv n^{\mu}n^{\nu}T_{\mu\nu}\,,\quad \quad p^{\mu}\equiv-f^{\mu\nu}n^{\rho}T_{\nu\rho}\,,
\quad \quad S^{\mu\nu}\equiv f^{\mu\rho}f^{\nu\sigma}T_{\rho\sigma}\,,
\label{eq:energy-momentum-def}
\end{equation}
where $E$ is the energy density, $p^{\mu}$ is the energy flux (or momentum) and $S^{\mu\nu}$ is stress tensor. Then, the standard energy-momentum tensor can be written as
\begin{equation}
T_{\mu\nu}=E n_{\mu}n_{\nu}+p_{\mu}n_{\nu}+p_{\nu}n_{\mu}+S_{\mu\nu}\,.\label{eq:EnergyMomentumTensor}
\end{equation}

Now we have everything that we need to present the decomposed Einstein equations. These are developed in full details in \cite{Mitsou:2020czr}, where the authors have specialized to the LC gauge as a $1+1+2$ ADM foliation. As a starting point, we can extract the energy (time-time) and momentum (time-space) constraints respectively given by\footnote{From now on, for a generic tensor $C_{ij}$, we will denote its trace with $C\equiv f^{ij}C_{ij}$.}
\begin{equation}
^{(3)}R+\Theta_{n}^{2}-K_{ij}K^{ij}=\,2E\,,
\qquad\qquad
-D_{j}K_{i}^{j}+D_{i}\Theta_{n}=\,p_{i}\,,\label{eq:energy-momentum-constraint}
\end{equation}
where we defined the extrinsic curvature as $K_{\mu\nu}\equiv\nabla_{(\mu}n_{\nu)}$. We can then also define the expansion rate $\Theta_{n}$ as
\begin{equation}
\Theta_{n}\equiv
f^{\mu\nu}K_{\mu\nu}\,.
\label{eq:expansionrate_def}
\end{equation}
The evolution of the induced metric $f_{ij}$ and of $K_{ij}$ is then obtained from the space-space decomposition
\begin{align}
\left(\partial_{t}-\mathcal{L}_{N}\right)f_{ij} &=\,2\mathcal{M}K_{ij}\,,\nonumber\\
\left(\partial_{t}-\mathcal{L}_{N}\right)K_{ij}&=\,\mathcal{M}\left[2K_{ik}K_{j}^{k}-K_{ij}\Theta_{n}-^{(3)}R_{ij}+S_{ij}-\frac{1}{2}f_{ij}\left(S-E\right)\right]+D_{i}D_{j}\mathcal{M}\,,\label{eq:curv-evolution}
\end{align}
where $\mathcal{L}_{N}$ is the Lie derivative along the field $N^i$.

Finally, the equations for the matter sector $\nabla_{\mu}T^{\mu\nu}=0$ are
given by
\begin{align}
\left(\partial_{t}-\mathcal{L}_{N}\right)E&=\,-D_{i}\left(\mathcal{M}p^{i}\right)-\mathcal{M}\left(\Theta_{n}E+K_{ij}S^{ij}\right)\,,\nonumber\\
\left(\partial_{t}-\mathcal{L}_{N}\right)p_{i}&=\,-D_{j}\left(S_{i}^{j}\mathcal{M}\right)-\mathcal{M}\Theta_{n}p_{i}-ED_{i}\mathcal{M}\,.\label{eq:momentum-evo}
\end{align}
Let us now follow the decomposition of \cite{Talebian-Ashkezari:2016llx} by extracting the shape-preserving volume expansion out of the spatial metric. First, we make a conformal re-scaling of $f_{ij}$ to 
\begin{equation}
f_{ij}\equiv e^{2\Xi}\hat{f}_{ij}\,,
\label{eq:conformal}
\end{equation}
by requiring that the determinant $\hat{f}=\text{det}[\hat{f}_{ij}]=1$. In this way, we can interpret $e^\Xi$ as the local effective scale factor. Then, we re-scale accordingly also the other quantities
\begin{align}
\hat{A}_{ij}= &\,e^{-2\Xi}\left(K_{ij} -\frac{1}{3}f_{ij}\Theta_{n}\right)\,,\nonumber \\
\,^{(3)}\hat{\mathcal{R}}_{ij}= &\,e^{-2\Xi}\left(^{(3)}R_{ij}-\frac{1}{3}\,^{(3)}R\,f_{ij}\right)\,,\nonumber \\
\hat{\mathcal{S}}_{ij}= &\,e^{-2\Xi}\left(S_{ij}- \frac{1}{3}f_{ij}S\right)\,.\label{eq:decomp}
\end{align}

Before applying the decomposition \eqref{eq:decomp} to Eqs.~\eqref{eq:curv-evolution} and \eqref{eq:momentum-evo},  we define $\mathcal{M}\frac{d}{d\widetilde{\lambda}}\equiv\left(\partial_{t}-\mathcal{L}_{N}\right)$ and compute the trace of the evolution of $A_{ij}\equiv e^{2\Xi}\hat{A}_{ij}$
\begin{align}
f^{ij}\left(\partial_{t}-\mathcal{L}_{N}\right)A_{ij}
= & \frac{1}{\mathcal{M}}\left[f^{ij}f_{il}\frac{d}{d\widetilde{\lambda}}\left(f_{jk}A^{lk}\right)+f^{ij}f_{jk}A^{lk}\frac{d}{d\widetilde{\lambda}}f_{il}\right]\nonumber \\
= & \frac{1}{\mathcal{M}}\frac{d}{d\widetilde{\lambda}}\left(f_{lk}A^{lk}\right)+f^{ij}f_{jk}A^{lk}2K_{il}\nonumber \\
= & 2A_{ij}A^{ij}=2\hat{A}_{ij}\hat{A}^{ij}\,,\label{eq:trace-less-evolution2}
\end{align}
where we have used the first of Eqs.~\eqref{eq:curv-evolution} from the first to the second line, and the fact that $A_{ij}$ is trace-less from the second to the last line.

Now we can apply the decomposition \eqref{eq:decomp} to Eqs.~\eqref{eq:curv-evolution}, using also Eqs. \eqref{eq:energy-momentum-constraint} and \eqref{eq:trace-less-evolution2}, to obtain
\begin{align}
\frac{d\Theta_{n}}{d\widetilde{\lambda}}= & -\frac{\Theta_{n}^{2}}{3}-\hat{A}_{ij}\hat{A}^{ij}-\frac{1}{2}\left(E+S\right)+\frac{1}{\mathcal{M}}D^{2}\mathcal{M}\,,\nonumber \\
\frac{d\Xi}{d\widetilde{\lambda}}= & \frac{\Theta_{n}}{3}\,,\nonumber \\
\frac{d\hat{f}_{ij}}{d\widetilde{\lambda}}= & 2\hat{A}_{ij}\,.\label{eq:evo-eqcs}
\end{align}
Then the evolution of the trace-less quantity $\hat{A}_{ij}$ is given by
\begin{equation}
\frac{d\hat{A}_{ij}}{d\widetilde{\lambda}}=-\frac{1}{3}\Theta_{n}\hat{A}_{ij}+2e^{-2\Xi}A_{ik}A_{j}^{k}+\hat{\mathcal{S}}_{ij}-\,^{(3)}\hat{\mathcal{R}}_{ij}+\frac{1}{\mathcal{M}}\left(D_{i}D_{j}-\frac{1}{3}f_{ij}D^{2}\right)\mathcal{M}\,.\label{eq:trace-less-evolution}
\end{equation}

At this point, we will apply our gradient expansion scheme without any gauge fixing. Hence, considering that $n^{\mu}=\frac{1}{\mathcal{M}}\left(1,-N^{i}\right)$, for a generic tensor $l_{ij}$, it holds
\begin{equation}
\frac{1}{\mathcal{M}}\left(\partial_{t}-\mathcal{L}_{N}\right)l_{ij}=n^{\mu}\partial_{\mu}l_{ij}+l_{ik}\partial_{j}N^{k}+l_{jk}\partial_{i}N^{k}=\frac{d}{d\lambda}l_{ij}+\mathcal{O}\left(\epsilon^{2}\right)\,,
\label{eq:generalevolution-shiftvector}
\end{equation}
where we define $n^{\mu}\partial_{\mu}\equiv\frac{d}{d\lambda}$ and then we have $\frac{d}{d\widetilde{\lambda}}=\frac{d}{d\lambda}+\mathcal{O}\left(\epsilon^{2}\right)$. This corresponds to neglect terms proportional to $\partial_{i}N^{j}=\mathcal{O}\left(\epsilon^{2}\right)$ in Eq.~\eqref{eq:generalevolution-shiftvector}, as done also in \cite{Sugiyama:2012tj,Talebian-Ashkezari:2016llx}. This is our main assumption and comes from the fact that, due to the spatial derivatives, the momentum constraint of Eq.~\eqref{eq:energy-momentum-constraint} is a first order relation, i.e. $p_{i}=\mathcal{O}\left(\epsilon\right)$. Hence, using Eq.~\eqref{eq:energy-momentum-def}, we get
\begin{align}
p_{i}=-\frac{1}{\mathcal{M}}\left(T_{it}
-N^{j}T_{ij}\right)\,.\label{eq:eq}
\end{align}
Such Eq.~\eqref{eq:eq} can be satisfied either by the strong condition
\begin{align}    T_{it}\sim N^{i}=\mathcal{O}\left(\epsilon\right)\,,
\label{eq:shiftgradcond}
\end{align}
or by the weaker condition that only the combination on its r.h.s. is $\mathcal{O}(\epsilon)$. Along this paper, we will adopt the stronger condition to justify our claim that $\pa_{i}N^j=\mathcal{O}(\epsilon^2)$, in agreement with \cite{Sugiyama:2012tj,Lyth:2004gb,Wands:2000dp,Talebian-Ashkezari:2016llx}.

Moreover, just as done in some previous
works \cite{Sugiyama:2012tj,Lyth:2004gb,Wands:2000dp},
we will also neglect $^{(3)}R_{ij}\thicksim R\thicksim\hat{\mathcal{S}}_{ij}=\mathcal{O}\left(\epsilon^{2}\right)$ since all of these terms contain double spatial derivatives. For what concerns the anisotropic stress $\hat{\mathcal{S}}_{ij}$, this is given by combinations of double spatial derivatives acting on the scalar fields in the matter sector.
The condition $\hat{\mathcal{S}}_{ij}=\mathcal{O}(\epsilon^{2})$ was relaxed in \cite{Talebian-Ashkezari:2016llx} only
for Bianchi geometries and on \cite{Tanaka:2021dww,Tanaka:2023gul} due to the presence of gauge fields.

With these considerations, we can decompose again the metric evolution provided by the first of Eqs.~\eqref{eq:curv-evolution} thanks Eqs.~\eqref{eq:decomp}. Hence, at first order in the gradient expansion, we obtain\footnote{Note that, although Eqs.~\eqref{eq:determinant-metric-evo} are very similar to Eqs.~\eqref{eq:evo-eqcs}, after the gradient expansion we have replaced $\frac{d}{d\tilde{\lambda}}=\frac{d}{d\lambda}+\mathcal{O}(\epsilon^{2})$.}
\begin{equation}
\frac{d\Xi}{d\lambda}=\frac{\Theta_{n}}{3}\,,\qquad\qquad\frac{d\hat{f}_{ij}}{d\lambda}=2\hat{A}_{ij}\,.\label{eq:determinant-metric-evo}
\end{equation}
Moreover, we can prove that $\hat{A}_{ij}$ is at least a second order term in the gradient expansion at every order in perturbation theory. In fact, following \cite{Sugiyama:2012tj}, from Eq.~\eqref{eq:trace-less-evolution} we get
\begin{equation}
\frac{d}{d\lambda}\hat{A}_{ij}=-\frac{1}{3}\Theta_{n}\hat{A}_{ij}+2\hat{A}_{ik}\hat{A}_{j}^{k}+\mathcal{O}\left(\epsilon^{2}\right)\,,
\label{eq:trace-less-evo-e2}
\end{equation}
and we choose a coordinate system such that $A_{ij}$ vanishes on the background\footnote{This is a quite general condition for isotropic spaces and, as we will show later, this is also the case for an isotropic LC background}. Hence, at $\mathcal{O}(\delta)$ in perturbation theory, we get that $\hat{A}_{ik}\hat{A}_{j}^{k}\sim \mathcal{O}(\delta^2)$ and then Eq.~\eqref{eq:trace-less-evo-e2}, with first of Eqs.~\eqref{eq:determinant-metric-evo}, becomes
\begin{equation}
\frac{d}{d\lambda}\hat{A}_{ij}=-\frac{d\Xi}{d\lambda}\hat{A}_{ij}+\mathcal{O}\left(\delta^{2},\epsilon^{2}\right)\,.
\label{eq:Aijdelta2}
\end{equation}
The latter equation is clearly solved by
\begin{equation}
\hat{A}_{ij}\propto e^{-\Xi}\,,
\label{eq:Aijdecaying}
\end{equation}
and proves that the $\mathcal{O}(1)$ term in the gradient expansion of $\hat{A}_{ij}$ decays when $\Xi$ grows in time. Then it can be neglected. Therefore, we obtain that $\hat{A}_{ij}$ is at least first order in $\epsilon$. As a consequence, $\hat{A}_{ik}\hat{A}_{j}^{k}$ is not only second order in $\delta$ expansion but also at least $\mathcal{O}\left( \epsilon^2 \right)$. Thanks to this result, this proof can be repeated iteratively to any order $n$-th in perturbation theory, leading then to
\begin{equation}
\frac{d}{d\lambda}\hat{A}_{ij}=-\frac{d\Xi}{d\lambda}\hat{A}_{ij}+\mathcal{O}\left(\delta^{n+1},\epsilon^{2}\right)\,.
\label{eq:Aijdelta-n}
\end{equation}
The solution in Eq.~\eqref{eq:Aijdecaying} solves also Eq.~\eqref{eq:Aijdelta-n} and this proves that also the term $\hat{A}_{ij}$, which is $\mathcal{O}(\epsilon)$, is decaying and can be neglected. So we have proven our initial claim that $\hat{A}_{ij}$ is at least of order $\epsilon^2$.
Hence, considering the evolution of
the spatial metric in Eqs.~\eqref{eq:determinant-metric-evo}, we have that
\begin{equation}
\frac{d}{d\lambda}\hat{f}_{ij}=
\mathcal{O}\left(\epsilon^{2}\right)\,,\label{eq:evolution-spatial-part}
\end{equation}
at any order in perturbation theory.

The above analysis was performed in \cite{Sugiyama:2012tj} leading also to $N^{i}=\mathcal{O}\left(\epsilon\right)$ when $N^{i}$ vanishes at the background. For the sake of clarity, we underline that in \cite{Sugiyama:2012tj} the UCG has been fixed and then $\hat{f}_{ij}$ corresponds to the tensor modes. This also shows that the evolution of the tensor modes can be neglected at linear order in $\epsilon$.

Finally, our complete set of equations to order $\mathcal{O}\left(\delta^{n+1},\epsilon^{2}\right)$
is given by the energy and momentum constraints
\begin{equation}
E=  \frac{\Theta_{n}^{2}}{3}\,,\hspace{3cm}
p_{i}= \frac{2}{3}D_{i}\Theta_{n}\,,\label{eq:generic-SUEqs}
\end{equation}
with their respective evolution equations given by
\begin{equation}
\frac{dE}{d\lambda}= -\Theta_{n}\left(E+\frac{1}{3}S\right)\,,\hspace{2cm}
\frac{dp_{i}}{d\lambda}= -\frac{1}{3\mathcal{M}}D_{i}\left(\mathcal{M}S\right)-\Theta_{n}p_{i}\,.\label{eq:generic-SUEqs2}
\end{equation}
In order to complete our set of SU equations we also need the decomposed spatial metric evolution given by Eqs. \eqref{eq:determinant-metric-evo} and \eqref{eq:evolution-spatial-part}. Also we have that the expansion rate evolution is given by
\begin{align} 
\frac{d\Theta_{n}}{d\lambda}= & -\frac{1}{3}\Theta_{n}^{2}-\frac{1}{2}\left(S+E\right)\,.\label{eq:generic-SUEqs3}
\end{align}
As one can easily see, Eqs.~\eqref{eq:determinant-metric-evo} and \eqref{eq:evolution-spatial-part}-\eqref{eq:generic-SUEqs3}, valid at first order in the gradient expansion and to all orders in perturbation theory, exactly correspond to the homogeneous and isotropic background equations if one neglects the momentum constraint. These then prove that the condition
\begin{equation}
\left(\partial_{t}-\mathcal{L}_{N}\right)f_{ij}=\frac{1}{\mathcal{M}}\frac{d}{d\lambda}f_{ij}+\mathcal{O}\left(\delta^{n+1},\,\epsilon^{2}\right)\,,
\end{equation}
with the fact that $\partial_{i}N^{i},\,\hat{\mathcal{S}}_{ij},\,R_{ij}$ and $R$ are $\mathcal{O}\left(\delta^{n+1},\,\epsilon^{2}\right)$,
reproduces the SU picture, and matches with previous works \cite{Sugiyama:2012tj,Lyth:2004gb,Wands:2000dp} if $\lambda= t$ and $N^{i}\partial_{i}=\mathcal{O}(\epsilon^{2})$.

In the next section, we will specialize this construction to non-linear LC perturbations on the top of a FLRW background. Thanks to the freedom of the choice of the lapse function in the LC gauge, we will then provide a general SU formalism. Furthermore, within the synchronous fixing of the lapse function, our formalism will be extended also to the GLC gauge.

\section{Light-Cone gauge}
\label{LCgauge}
Let us now introduce the LC gauge \cite{Mitsou:2020czr}. This is a generalization of the GLC gauge \cite{Gasperini:2011us}, where the lapse function is left unfixed and is built as a foliation of the spacetime thanks to a set of four coordinates adapted to the observed past light-cone. In particular, the proper time of a generic observer is described by the coordinate $t$. This corresponds to the proper-time of a free-falling observer when the GLC fixing of the lapse function occurs. The coordinates $w$ and $\theta^{a}$ satisfy the same properties as in the GLC gauge, i.e. $w$ describes the observer's past light-cone and $\theta^a=const$ describes the light-like geodesics. Given that, the non-linear line element is \cite{Mitsou:2020czr}
\begin{equation}
ds_{LC}^{2}=\Upsilon^{2}dw^{2}-2\mathcal{M}\Upsilon dwdt+\gamma_{ab}\left(d\theta^{a}-U^{a}dw\right)\left(d\theta^{b}-U^{b}dw\right)\,.\label{eq:LCgauge}
\end{equation}

In this case, the vector $n^\mu$ in Eq.~\eqref{eq:def-n} is given by
\begin{equation}
n^{\mu}=\left(\frac{1}{\mathcal{M}},\,\frac{1}{\Upsilon},\,\frac{U^{a}}{\Upsilon}\right)\,.
\label{eq:orthog}
\end{equation}

The advantage of the metric \eqref{eq:LCgauge} is that it simplifies the description of light-like signal. For instance, the light-like geodesics are exactly solved by $k_{\mu}=-\omega\delta_{\mu}^{w}$, where $k_{\mu}$ is the four-momentum of the photon and $\omega$ is its physical frequency. Moreover, for the GLC gauge where $\mathcal{M}=1$, also the time-like geodesic is exactly solved by $u_{\mu}=-\partial_{\mu}t$. In this case $u_{\mu}=n_{\mu}$ is the four-velocity of the geodesic observer and is perpendicular to the three-dimensional hypersurfaces of $t=const$. This particular choice simplifies the description of late-time cosmological observables and allows a completely non-linear description of such observables as a factorization of the metric entries \cite{Gasperini:2011us,BenDayan:2012pp,BenDayan:2012ct,BenDayan:2012wi,Fanizza:2013doa,Marozzi:2014kua,Fanizza:2019pfp,Fanizza:2020xtv,DiDio:2014lka,DiDio:2015bua,Marozzi:2016uob,Marozzi:2016qxl}. As relevant examples, the expressions for the cosmological redshift  and the angular distance are given, in a exact way and for an arbitrary geometry, directly as \cite{Gasperini:2011us,Fanizza:2013doa}
\begin{equation}
1+z=\frac{\left(u_{\mu}k^{\mu}\right)_{s}}{\left(u_{\nu}k^{\nu}\right)_{o}}=\frac{\Upsilon_{o}}{\Upsilon_{s}}\,,\qquad\qquad d^{2}_{A}=\frac{\sqrt{\gamma}}{\left(\frac{\text{det}\partial_{\tau}\gamma_{ab}}{4\sqrt{\gamma}}\right)_{o}}\,,\label{eq:observables}
\end{equation}
where $z$ is the redshift of the source, the subscript $s$ and $o$ stands for a quantity evaluated at the source and observer position, and $\gamma$ is the determinant of $\gamma_{ab}$.

\subsection{LC gauge shift vector}
Here, we will use $N_{LC}^{i}$ (instead of $N^{i}$)\footnote{For the LC metric, Latin indices will always refer to the coordinates $w$ and $\theta^a$.} to describe the shift vector of the LC gauge, thus avoiding confusion when we relate $N_{LC}^{i}$ to the standard $N^{i}$.
Hereafter, we will provide a SU picture allowing for different lapse function fixings within the LC gauge \cite{Mitsou:2020czr}. This is a crucial step to obtain the $\delta\mathcal{N}$ formalism on the past light-cone.

In \cite{Mitsou:2020czr} was shown that the non-linear LC gauge
can be interpreted as a $1+1+2$ ADM decomposition with coordinates
$x^{\mu}=\left(t,w,\theta^{a}\right)$. This proviso, the shift vector for the first $1+3$ decomposition is then given by
\begin{equation}
N_{LC}^{i}=-\mathcal{M}\left(\frac{1}{\Upsilon},\frac{U^{a}}{\Upsilon}\right)\,,\qquad\qquad N^{LC}_{j}=-\Upsilon\delta_{j}^{w}\,.\label{eq:GLC-shiftvector}
\end{equation}

For what concerns the shift vector $N^{i}_{LC}$, this is orthogonal to the surface at constant $t$ and $w$. Hence, if we recall that the photon four-momentum in the LC coordinates is $k^\mu=\omega\mathcal{M}^{-1}\Upsilon^{-1}\delta^\mu_t$, within the 1+3 decomposition we can write the shift vector as 
\begin{equation}
N^i_{LC}=\frac{k^i}{\omega}-n^i\,.
\end{equation}
Hence, since $n^\mu$ is a time-like vector, $N^i_{LC}$ can be interpreted as the space-like component of the propagation direction of an incoming photon (see \cite{Fleury:2016htl}). To completely fix the LC gauge, we still have to fix three conditions. These are given by the following ones
\begin{equation}
f_{ww}=\Upsilon^{2}+U^{2}\,,\qquad\qquad
f_{wa}=-U_{a}\,,\qquad\qquad
f_{ab}=\gamma_{ab}\,.\label{eq:GLC-ADM-conditions}
\end{equation}

As one can see, $N_{LC}^{w}=\frac{1}{a}$ on the background level, so it does not vanish when $\epsilon\rightarrow0$. We then have that $N_{LC}^{w}=\mathcal{O}\left(1\right)$.
With this choice of coordinates, however, we will show that the condition $\partial_{i}N_{LC}^{i}=\mathcal{O}\left(\epsilon^{2}\right)$ holds.  In order to do so, we first perform a finite background coordinate transformation on the metric in Eq. \eqref{eq:ADM1}, from the $x^{i}=\left(r,\theta^{a}\right)$
coordinates to the light-cone ones $y^{i}=\left(w,\theta^{a}\right)$, given by
\begin{equation}
dr=dw-\frac{dt}{a}\,,\qquad\qquad
d\theta^{a}=d\theta^{a}\,,\qquad\qquad
dt=dt\,,\label{eq:finite_coordinate}
\end{equation}
in order to relate $N^i$ to $N_{LC}^i$. A direct computation for the controvariant components of $N^i$ returns that
\begin{equation}
N^{r}= \frac{1}{a}-\frac{\mathcal{M}}{\Upsilon}\,,
\qquad\qquad
N^{a}=-\frac{U^{a}}{a}\left(1-\frac{1}{a}+\frac{\mathcal{M}}{\Upsilon}\right)\,,\label{eq:beta-covariant}
\end{equation}
or, equivalently, in a covariant form
\begin{equation}
N_{r}=-\mathcal{M}\Upsilon+\frac{\Upsilon^{2}+U^{2}}{a}\,,\label{eq:beta-r}
\end{equation}
and
\begin{equation}
N_{a}=-\frac{U_{a}}{a}\,.\label{eq:beta-a}
\end{equation}

Finally, using Eqs.~\eqref{eq:GLC-shiftvector} and \eqref{eq:beta-covariant}, the gradient expansion condition of Eq.~\eqref{eq:shiftgradcond} given by $N^{i}=\mathcal{O}(\epsilon)$ returns $\partial_{i}N^{i}=\partial_{i}N_{LC}^{i}=\mathcal{O}\left(\epsilon^{2}\right)$. Within the gradient expansion, we then have
\begin{align}
\partial_{r}N^{r}= & -\partial_{w}\left(\frac{\mathcal{M}}{\Upsilon}\right)=\mathcal{O}\left(\epsilon^{2}\right)\,,\nonumber \\
\partial_{a}N^{a}= &  -\frac{1}{a}\left(1-N^{r}\right)\partial_{a}U^{a}+U^{a}\partial_{a}N^{r}
=\mathcal{O}\left(\epsilon^{2}\right)\,,\label{eq:gradexp}
\end{align}
which indeed show that $\partial_{w}\Upsilon^{-1}\sim\partial_{w}\mathcal{M}\sim\partial_{a}U^{a}=\mathcal{O}\left(\epsilon^{2}\right)$. This comes from the fact that both $\mathcal{M}$ and $\Upsilon$ have background counterparts, therefore, both $\partial_{w}\Upsilon^{-1}$ and $a^{-1}\partial_{w}\mathcal{M}$ are of order $\epsilon^{2}$. The condition $\partial_{a}U^{a}=\mathcal{O}(\epsilon^{2})$  is obtained by using the fact that $N^{r}=\mathcal{O}(\epsilon)$ on the second of Eqs.~\eqref{eq:gradexp}.

\subsection{Separate Light-Cones}
\label{SGLCU}
As it has been shown in Sect.~\ref{SADMU}, one can still obtain a SU picture when the shift vector combines to form an integral along the geodesics and only its spatial derivatives are neglected. We will prove that this is the case for $\partial_{i}N^{i}$ in Eq.~\eqref{eq:shiftgradcond} and for $\partial_{i}N_{LC}^{i}$ in Eq.~\eqref{eq:gradexp}. Additionally, we need the trace-less part of the extrinsic curvature $\hat{A}_{ij}$ to be negligible in order to obtain the SU scheme for the LC metric.

The condition for the shift vector is given by
\begin{equation}
\partial_{i}N_{LC}^{i}=\partial_{w}\left(\frac{\mathcal{M}}{\Upsilon}\right)+\partial_{a}\left(\frac{\mathcal{M}U^{a}}{\Upsilon}\right)=\mathcal{O}\left(\epsilon^{2}\right)\,,\label{eq:shift-v-GLC-cond}
\end{equation}
which, with Eqs.~\eqref{eq:curv-evolution}, implies
\begin{equation}
\frac{d}{d\lambda}f_{ij}=2K_{ij}+\mathcal{O}\left(\epsilon^{2}\right)\,,\label{eq:GLC-extrinsic curv}
\end{equation}
where in $\mathcal{L}_{N}f_{ij}$ we have neglected $\partial_{i}N^{i}$ but not $N^{i}\partial_{i}f_{jk}$. In fact, the latter combines with $\partial_{t}f_{ij}$ to reconstruct $\frac{d}{d\lambda}f_{ij}$, following the general prescription given in Eq.~\eqref{eq:generalevolution-shiftvector}. One may note the similarity between Eqs.~\eqref{eq:GLC-extrinsic curv} and \eqref{eq:curv-evolution}. This is because Eq.~\eqref{eq:curv-evolution} is the non-perturbative version in the gradient expansion of Eq.~\eqref{eq:GLC-extrinsic curv}, where we consider instead the gradient expansion on the parameter $\frac{d}{d\tilde{\lambda}}=\frac{d}{d\lambda}+\mathcal{O}(\epsilon^{2})$.

Using Eqs.~\eqref{eq:GLC-ADM-conditions} at the background level, i.e.
\begin{equation}
f_{ww}=a^{2}\,,\qquad\qquad
f_{wa}=0\,,\qquad\qquad
f_{ab}=a^{2}r^{2}\bar{q}_{ab}\,,
\end{equation}
one gets that $d f_{ij}/d\lambda =2H f_{ij}$, where $H$ is the background expansion rate defined as $H\equiv \bar{\Theta}_n/3$, $\bar{\Theta}_n$ is the extrinsic curvature on the background and $\bar{q}_{ab}=\text{diag}\left( 1,\sin\theta \right)$. Thus, from Eq.~\eqref{eq:GLC-extrinsic curv}, we also see that $\hat{A}_{ij}$ vanishes on the background. Following the same procedure adopted in Eqs.~\eqref{eq:Aijdelta2} and \eqref{eq:Aijdelta-n}, we get that $\hat{A}_{ij}=\mathcal{O}\left(\epsilon^{2}\right)$
also when the LC gauge is fixed.

Now, by taking the trace of Eq.~\eqref{eq:GLC-extrinsic curv}, we obtain
\begin{align}
\Theta_{n}&=\,\frac{1}{\sqrt{-g}}\partial_{\mu}\left(\sqrt{-g}n^{\mu}\right)\nonumber\\
&=\,\frac{1}{\mathcal{M}\Upsilon\sqrt{\gamma}}\frac{d}{d\lambda}\left(\mathcal{M}\Upsilon\sqrt{\gamma}\right)+\partial_{\mu}n^{\mu}\nonumber\\
&=\,\frac{1}{\Upsilon\sqrt{\gamma}}\frac{d}{d\lambda}\left(\Upsilon\sqrt{\gamma}\right)+\frac{1}{\mathcal{M}}\frac{d}{d\lambda}\mathcal{M}+\partial_{\mu}n^{\mu}\nonumber\\
&=\,\frac{1}{\Upsilon\sqrt{\gamma}}\frac{d}{d\lambda}\left(\Upsilon\sqrt{\gamma}\right)+\frac{1}{\mathcal{M}}\left[\partial_{w}\left(\frac{\mathcal{M}}{\Upsilon}\right)+\partial_{a}\left(\frac{\mathcal{M}U^{a}}{\Upsilon}\right)\right]\nonumber\\
&=\,\frac{1}{\Upsilon\sqrt{\gamma}}\frac{d}{d\lambda}\left(\Upsilon\sqrt{\gamma}\right)-\frac{1}{\mathcal{M}}\partial_{i}N_{LC}^{i}\,.\label{eq:K-hij-trace}
\end{align}
Thanks to this last equation, within the fixing $\mathcal{M}=1$, we realize that the difference between $\Theta_{u}\equiv\nabla_\mu u^\mu$ and $\Theta_{n}$ is of order $\epsilon^{2}$. We also have from Eq.~\eqref{eq:determinant-metric-evo}, where we neglect $\partial_{i}N^{i}_{LC}$ in $\Theta_{n}$
\begin{equation}
\frac{d\,\Xi}{d\lambda}=\frac{1}{3\Upsilon\sqrt{\gamma}}\frac{d\left(\Upsilon\sqrt{\gamma}\right)}{d\lambda}+\mathcal{O}\left(\epsilon^{2}\right)\,,
\label{perturbedefold-trK}
\end{equation}
which preserves the background form.

Considering now the homogeneous and isotropic LC background, namely $\bar{\Upsilon}\sqrt{\bar{\gamma}}=a^{3}r^{2}$, where the bar refers to background quantities,
Eqs.~\eqref{eq:generic-SUEqs} return
\begin{align}
&\frac{\bar{\Theta}_{n}^{2}}{3}=3H^{2}=3\left(\frac{\partial_{t}a}{a}\right)^{2}=\bar{E}\,,\hspace{6cm}\text{background}
\nonumber\\
&\frac{{\Theta_{n}}^{2}}{3}=\frac{1}{3}\left[\frac{1}{(\Upsilon\sqrt{\gamma})}\frac{d\left(\Upsilon\sqrt{\gamma}\right)}{d\lambda}\right]^{2}=E+\mathcal{O}\left(\delta^n,\epsilon^{2}\right)\,.\hspace{2cm}\text{non-perturbative}
\label{eq:H-constraint}
\end{align}
Moreover, Eqs.~\eqref{eq:generic-SUEqs2} and~\eqref{eq:generic-SUEqs3} give at the background level
\begin{equation}
\partial_{t}\bar{E}= \,-3H\left(\bar{E}+\frac{
1
}{3}\bar{S}\right)\,,\qquad\qquad
\partial_{t}\bar{\Xi}= \,H\,,\qquad\qquad
\partial_{t}H= \,-3H^{2}-\frac{1}{2}\left(\bar{E}+\bar{S}\right)\,.\label{eq:second-Friedmann-Eq}
\end{equation}
Here we remark that Eqs.~\eqref{eq:H-constraint} mean that the non-linear LC perturbations in a FLRW universe at first order in the gradient expansion do evolve as a set of glued background universes with a different set of $\left(a,\,E,\,S\right)$ in each patch. Within this picture, then $\Upsilon\sqrt{\gamma}$ can be linked to the effective local scale factor. To complete the set of equations that have the same form for the background (last of Eq.~\eqref{eq:second-Friedmann-Eq}) and the perturbed universe, we also have from Eq.~\eqref{eq:generic-SUEqs3}
\begin{equation}
\frac{d}{d\lambda}\left[\frac{1}{(\Upsilon\sqrt{\gamma})}\frac{d(\Upsilon\sqrt{\gamma})}{d\lambda}\right]=-\frac{1}{3} \left[\frac{1}{(\Upsilon\sqrt{\gamma})}\frac{d(\Upsilon\sqrt{\gamma})}{d\lambda}\right]^2-\frac{1}{2}\left(S+E\right)+\mathcal{O}\left(\delta^{n},\epsilon^{2}\right)\,.\label{eq:extrinsic-curv-evo-GLC}
\end{equation}
Therefore, the SU scheme holds with the universe evolving as a set of homogeneous and isotropic LC background, where $\lambda$ provides the evolution of inhomogeneities along $n^{\mu}$. So far we have provided a consistent SU on the past light-cone in terms of LC gauge entries, which is a fundamental step to provide the $\delta\mathcal{N}$ formalism on the past light-cone in the next Sect.~\ref{deltaN}.

\subsection{The Geodesic Light-Cone gauge}
\label{GLCgauge}
Let us now apply the SU scheme described in the previous subsection to the case of the GLC gauge given by Eq.~\eqref{eq:LCgauge} with $\mathcal{M}=1$, and show how it simplifies the evolution of the density perturbations on the past light-cone. This gauge automatically provides $\nabla_{\mu}u^{\mu}$=$\nabla_{\mu}n^{\mu}$,
then, one can relate the expansion of the 3D-hypersurfaces orthogonal to $n^{\mu}$ to the matter content described in terms of $u^{\mu}$.

Hence, we recall that the comoving four velocity is given by $u^{\mu}=\left(1,\Upsilon^{-1},\Upsilon^{-1}U^{a}\right)$ 
\cite{Fleury:2016htl,Gasperini:2011us}, we then have that the expansion of the 3D-hypersurfaces is
\begin{equation}
\Theta_{u}
=\nabla_{\mu}u^{\mu}=\frac{\partial_{t}\Upsilon}{\Upsilon}+\frac{\gamma^{ab}\partial_{t}\gamma_{ab}}{2}+\frac{\gamma^{ab}}{2\Upsilon}\partial_{w}\gamma_{ab}+\frac{1}{\Upsilon}\partial_{a}U^{a}
+\frac{U^{a}\gamma^{bc}}{2\Upsilon}\partial_{a}\gamma_{bc}\,,
 \label{eq:expansion-non-linear}
\end{equation}
which can be re-written in a more suitable form as
\begin{equation}
\Theta_{u}=\frac{1}{\Upsilon\sqrt{\gamma}}\frac{d\left(\Upsilon\sqrt{\gamma}\right)}{d\lambda}
+\partial_\mu u^\mu\,,
\label{eq:expansion_3}
\end{equation}
where $\frac{d}{d\lambda}\equiv u^{\mu}\partial_{\mu}$ accounts for inhomogeneities along the geodesics and $\partial_{\mu}u^{\mu}=\partial_{i}N^{i}$ is $\mathcal{O}(\epsilon^2)$, as shown in Eqs.~\eqref{eq:gradexp}.

An interesting feature of Eq.~\eqref{eq:expansion_3}
is that the first term contributes both to the background and to the perturbative level whereas the last term contributes only
to the perturbative level. We thus obtain a separate universe description using Eq. \eqref{eq:expansion_3} as aimed.

In order to provide the conservation of $\zeta$, when the pressure is adiabatic, we need to analyze the energy-momentum conservation in the GLC gauge for the case of a perfect fluid. Starting from
\begin{equation}
T_{\mu\nu}=\left(\rho+p\right)u_{\mu}u_{\nu}+g_{\mu\nu}p\,,
\label{eq:perfect_fluid}
\end{equation}
where $\rho$ and $p$ respectively describes the energy-density and pressure as measured by an free-falling observer. The conservation law along the direction of $u^{\nu}$, i.e. $u^{\nu}\nabla_{\mu}T_{\,\nu}^{\mu}  =0$, exactly returns
\begin{equation}
\frac{d\rho}{d\lambda}=-\left(\rho+p\right)\Theta_{u}\,.
\label{eq:conserv}
\end{equation}
Hence, by using Eq.~\eqref{eq:expansion_3}, we have
\begin{align}
\frac{d\rho}{d\lambda}=-\left(\rho+p\right)\left[\frac{1}{\Upsilon\sqrt{\gamma}}\frac{d\left( \Upsilon\sqrt{\gamma} \right)}{d\lambda}
+\partial_\mu u^\mu\right]
\,.\label{eq:conservation}
\end{align}
Eq.~\eqref{eq:conservation} is a fully non-linear relation between geometry and matter content.
As an important remark, since Eq.~\eqref{eq:conservation} is a fully non-linear equation, it can be seen as a dynamic equation for the exact density perturbations.

According to what outlined so far, within the gradient expansion, where $\partial_{\mu}u^{\mu}=\partial_{i}N^{i}=\mathcal{O}\left(\epsilon^{2}\right)$, Eq.~\eqref{eq:conservation} can then be written as
\begin{align}
\frac{d\rho}{d\lambda}
=&-\left(\rho+p\right)\frac{1}{\Upsilon\sqrt{\gamma}}\frac{d\left( \Upsilon\sqrt{\gamma} \right)}{d\lambda}+\mathcal{O}(\epsilon^2)\,.
\label{eq:conservation_gradient}
\end{align}
In general, pressure and energy density are linked by an equation of state as $p =q\rho$. The value of $q$ can be time-dependent, accordingly to the specific era when inhomogeneities are evolving (as happens, for instance, during the slow-roll inflationary stage). This makes Eq.~\eqref{eq:conservation_gradient} in general quite complicated to be solved. However, during the late time epochs (e.g. radiation, matter  or cosmological constant dominated universe), $q$ is constant. In this case, Eqs.~\eqref{eq:conservation_gradient} becomes
\begin{align}
\frac{d\rho}{d\lambda}
=&-\rho\frac{1+q}{\Upsilon\sqrt{\gamma}}\frac{d\left( \Upsilon\sqrt{\gamma} \right)}{d\lambda}+\mathcal{O}(\epsilon^2)\,,
\label{eq:conservation_gradient_perfect}
\end{align}
which is exactly solved by
\begin{equation}
\rho(\lambda)= A \left( \Upsilon\sqrt{\gamma} \right)^{-(1+q)}(\lambda)
+\mathcal{O}(\epsilon^2)\,,
\label{eq:exact_rho}
\end{equation}
where $A$ is a constant. Eq.~\eqref{eq:exact_rho} gives the exact link between the geometry and the energy density in terms of light-cone metric entries. It is also the starting point to describe non-linear features of inhomogeneities on super-horizon scales. In fact, in the following we will discuss how $\Upsilon\sqrt{\gamma}$ relates with the gauge invariant curvature perturbation $\zeta$ and how both can be computed using the $\delta\mathcal{N}$ formalism. Furthermore, we will show under which conditions the quantity $\zeta$ is conserved on super-horizon scales.

\section{Curvature perturbation evolution}
\label{deltaN}
In this section we will study the curvature perturbations along the past light-cone using the SU scheme developed in the previous section. To this aim, we will start by considering the linear order in perturbation theory, and by using the scalar-pseudoscalar decomposition developed in \cite{Fanizza:2020xtv, Fanizza:2023ixk}. Hence, we will first review this perturbation theory, following the approach of \cite{Wands:2000dp, Lyth:2004gb} to show the conservation of $\zeta$ along the past light-cone. Thereafter, we will generalize this proof to non-linear order in the amplitude of the perturbations and to first order in the gradient expansion, which allow us to obtain $\zeta$ at this perturbative level in terms of light-cone perturbations. Finally, we will generalize the $\delta\mathcal{N}$ formalism on the past light-cone.

\subsection{Linear evolution and comparison with previous results}
\label{linearevo}
In this section, we want to linearize Eqs.~\eqref{eq:conservation_gradient_perfect}. To this aim, we first recall the linear perturbation theory for the GLC coordinates, presented in \cite{Fanizza:2020xtv,Fanizza:2023ixk}. Firstly, we consider general perturbations, i.e. without fixing the GLC gauge. The metric and its perturbed inverse are then given by
\begin{equation}
g_{\mu\nu}=
\bar{g}^{\,GLC}_{\mu\nu}+\delta g_{\mu\nu}=a^{2}\left\{ \begin{pmatrix}0 & -a^{-1} & \vec{0}\\
-a^{-1} & 1 & \vec{0}\\
\vec{0}^{\mathbf{T}} & \vec{0}^{\mathbf{T}} & \bar{\gamma}_{ab}
\end{pmatrix}+\begin{pmatrix}L & M & V_{b}\\
M & N & \Ucal_{b}\\
V_{a}^{\mathbf{T}} & \Ucal_{a}^{\mathbf{T}} & \delta\gamma_{ab}
\end{pmatrix}\right\} \,,\label{eq:gen_pert_GLC_metric}
\end{equation}
and
\begin{equation}
\delta g^{\mu\nu}=\begin{pmatrix}-\left(a^{2}L+N+2aM\right) & -a^{-1}\left(a^{2}L+aM\right) & -a\left(aV^{a}+\Ucal^{a}\right)\\
-a^{-1}\left(a^{2}L+aM\right) & -L & aV^{a}\\
-a\left(aV^{a}+\Ucal^{a}\right) & aV^{a} & -a^{-2}\delta\gamma^{ab}
\end{pmatrix}\,,
\end{equation}
where, following \cite{Fanizza:2020xtv,Fanizza:2023ixk}, we can decompose $\Ucal_{a}, V_{a}$ and $\delta\gamma_{ab}$ as
\begin{align}
\Ucal_{a}&=r^{2}\left(D_{a}u+\tilde{D}_{a}\hat{u}\right)\,,\nonumber\\
V_{a}&=r^{2}\left(D_{a}v+\tilde{D}_{a}\hat{v}\right)\,,\nonumber\\
\delta\gamma_{ab}&=a^{2}r^{2}\left[\left(1+2\nu\right)\bar{q}_{ab}+D_{ab}\mu+\tilde{D}_{ab}\hat{\mu}\right] \,.\label{eq:scalar-pseudo-scalar}
\end{align}
Here, $u,\,v,\,\nu$ and $\mu$ are scalars, and $\hat{u},\,\hat{v}$ and $\hat\mu$ are pseudoscalar degrees of freedom under spatial rotations. Moreover, the angular derivatives are defined as 
\begin{equation}
D_{ab}=D_{(a}D_{b)}-\frac{1}{2}q_{ab}D^{2}\,,\qquad\qquad \tilde{D}_{ab}=D_{(a}\tilde{D}_{b)}\,,
\label{eq:Dab}
\end{equation}
where $\tilde{D}_{a}=\epsilon_{a}^{b}D_{b}$, and $\epsilon_{a}^{b}$ is the anti-symmetric tensor. We remark that $D_{ab}$ and $\tilde{D}_{ab}$ are trace-less, so that the trace of $\delta\gamma_{ab}$ is given by the trace of $q_{ab}=\left(1+2\nu\right)\bar{q}_{ab}$.

At this point, let us also linearize the energy density as
\begin{equation}
\rho=\bar\rho\left(1+\delta\rho\right)\,.\label{eq:expansion}
\end{equation}
In this expansion, however, we keep both $\bar\rho$ and $\delta\rho$ as function of the time-like parameter $\lambda$. By doing this, we then keep track of the perturbations as projected onto the exact time-like geodesic. Starting from Eqs.~\eqref{eq:scalar-pseudo-scalar}, we then obtain that
\begin{equation}
\sqrt{\gamma}=a^2\sqrt{\bar\gamma}\,\left(1+2\nu\right)\,.
\end{equation}


At this point, we want to use the metric in Eq. \eqref{eq:gen_pert_GLC_metric} to compute the expansion volume $\Theta_n=\nabla_{\mu}n^{\mu}$ 
of the hypersurfaces orthogonal to $t$ defined in Eq.~\eqref{eq:def-n}. Then, the volume expansion will be given by
\begin{equation}
\Theta_{n}=\frac{1}{\sqrt{-g}}\partial_{\mu}\left(\sqrt{-g}\,n^{\mu}\right)=\frac{1}{2}g^{\alpha\beta}\frac{d}{d\lambda}g_{\alpha\beta}+\partial_{\mu}n^{\mu}\,,\label{theta-GLC1}
\end{equation}
where $n^{\mu}\partial_{\mu}=\frac{d}{d\lambda}$.
Using this last equation, and the fact that  $\Theta_{n}=\Theta_{u}+\mathcal{O}(\epsilon^{2})$, as shown in Eq.~\eqref{eq:K-hij-trace}, we then have
\begin{equation}
\Theta_{u}=3H\left[1+\frac{1}{2}\left(a^{2}L+N+2aM\right)\right]+\frac{1}{2}\frac{d}{d\lambda}\left(N+4\nu\right)+\mathcal{O}(\epsilon^{2})\,\label{eq:linear-theta},
\end{equation}
where we are neglecting the terms
\begin{equation}
\left(\partial_{w}+\frac{2}{r}\right)\left(N+aM\right)=\mathcal{O}\left(\epsilon^{2}\right)\,,
\qquad\qquad D_{a}\left(\Ucal^{a}+aV^{a}\right)=\mathcal{O}\left(\epsilon^{2}\right)\,,
\end{equation}
since they are proportional to $\nabla_{i}\mathcal{B}^{i}$ in standard
perturbation theory (see \cite{Fanizza:2020xtv,Fanizza:2023ixk}). The same happens in the standard approach, see \cite{Wands:2000dp, Lyth:2004gb}. The reason why $\nabla_{i}\mathcal{B}^{i}$ can be neglect is that, in the language of the gradient expansion applied to the ADM metric, this term is the divergence of the shift vector, hence second order in the gradient expansion. As a side remark, we underline that Eq.~\eqref{theta-GLC1} is given in terms of the cosmic time by
\begin{equation}
\Theta_{n}=3H-3\partial_{t}\psi+\mathcal{O}(\epsilon^{2})\,,\label{eq:linear-theta-SVT}
\end{equation}
which is in agreement with \cite{Wands:2000dp}.

Let us now consider the energy-momentum tensor conservation projected onto $u^{\mu}$. From Eq.~\eqref{eq:conserv}, we have that
\begin{align}
\frac{d\rho}{d\lambda}+\left(\rho+p\right)\Theta_{u}=0\,.
\label{eq:linear-density-evo}
\end{align}
Thanks to Eqs.~\eqref{eq:expansion} and \eqref{eq:linear-theta}, and by using the equation of state $p=q \rho$, Eq.~\eqref{eq:linear-density-evo} gives 
\begin{align}
&\frac{d\bar\rho}{d\lambda}+3H\bar\rho (1+q)=0\,,
\nonumber\\
&\frac{d\left( \bar\rho\delta\rho \right)}{d\lambda}
+\bar\rho\left(1+q\right)\left\{ 3H\left[\delta\rho+\frac{1}{2}\left(a^{2}L+N+2aM\right)\right]+\frac{1}{2}\frac{d}{d\lambda}\left(N+4\nu\right)\right\}+\mathcal{O}(\epsilon^{2}) =0\,,
\label{eq:linear-nongauge-fixed}
\end{align}
for the background and perturbed quantities respectively.
Let us now fix the GLC gauge, which satisfies the following conditions \cite{Fanizza:2020xtv}
\begin{align}
a^{2}L+N+2aM&=0\,,\nonumber\\
\partial_{w}\left(N+aM\right)&=\frac{1}{2}\partial_{w}N=\mathcal{O}(\epsilon^2)\,,\nonumber\\
\partial_{a}\left(\Ucal^{a}+aV^{a}\right)&=\partial_{a}\Ucal^{a}=\mathcal{O}\left(\epsilon^{2}\right)\,.
\label{eq:GLCgaugecond}
\end{align}
Then, by inserting first of Eqs.~\eqref{eq:linear-nongauge-fixed} into the second one, by using the GLC gauge conditions given in Eqs.~\eqref{eq:GLCgaugecond}, and, finally, by integrating over the affine parameter $\lambda$, we get
\begin{equation}
\delta\rho(\lambda)=-\frac{ 1+q}{2}\left(N+4\nu\right)\,.
\end{equation}
It is worth to stress that we would obtain the same result by linearizing  Eqs.~\eqref{eq:exact_rho}. Therefore, this show the self-consistency of our gradient expansion method.

At first order in the gradient expansion, for a generic adiabatic equation of state, i.e. $p=p(\rho)$, and by fixing the GLC gauge, we obtain, from Eqs.~\eqref{eq:linear-theta}, \eqref{eq:linear-density-evo} and \eqref{eq:GLCgaugecond}, that
\begin{equation}
\frac{1}{\rho+p(\rho)}\frac{d\rho}{d\lambda}=-3H-\frac{1}{2}\frac{d\left(N+4\nu\right)}{d\lambda}\,.\label{eq:fluid-geometry}
\end{equation}
This equation can be integrated between two different hypersurfaces marked by the values $\lambda_1$ and $\lambda_2$, and gives
\begin{equation}
\frac{1}{6}\left(N+4\nu\right)|_{\lambda_{1}}^{\lambda_{2}}+\Nbcal(\lambda_{2},\lambda_{1})=-\frac{1}{3}\int_{\rho(\lambda_{1},x^{i})}^{\rho(\lambda_{2},x^{i})}\frac{d\rho}{\rho+p(\rho)}\,,\label{eq:integrated}
\end{equation}
where the dependence on $x^i$ in the extremes of integration denotes that we are considering inhomogeneous hyper-surfaces. At the same time, we have defined $\bar{\mathcal{N}}$ as
\begin{align}
\bar{\mathcal{N}}\left(\lambda_{2},\,\lambda_{1}\right)\equiv\int^{\lambda_{2}}_{\lambda_{1}} Hd\lambda\label{eq:back-e-folds}
\end{align}
Moreover, as done for standard perturbations in \cite{Lyth:2004gb}, we want to extract the background contribution of Eq.~\eqref{eq:integrated}. Therefore, we start by doing this in the r.h.s. of Eq.~\eqref{eq:integrated}, which can then be written as
\begin{equation}
\int_{\rho(\lambda_{1},x^{i})}^{\rho(\lambda_{2},x^{i})}\frac{d\rho}{\rho+p(\rho)}=\int_{\bar{\rho}(\lambda_{2})}^{\rho(\lambda_{2},x^{i})}\frac{d\rho}{\rho+p(\rho)}-\int_{\bar{\rho}(\lambda_{1})}^{\rho(\lambda_{1},x^{i})}\frac{d\rho}{\rho+p(\rho)}+\int_{\bar{\rho}(\lambda_{1})}^{\bar{\rho}(\lambda_{2})}\frac{d\rho}{\rho+p(\rho)}\,,\label{eq:procedure}
\end{equation}
where $\bar{\rho}$ determines the background value of $\rho$. The last
term is the background value given by $\Nbcal$ (see Eq.~\eqref{eq:back-e-folds}). Therefore, we have
\begin{equation}
\frac{1}{6}\left(N+4\nu\right)\left(\lambda_{1}\right)+\frac{1}{3}\int_{\bar{\rho}(\lambda_{1})}^{\rho(\lambda_{1},x^{i})}\frac{d\rho}{\rho+p(\rho)}=\frac{1}{6}\left(N+4\nu\right)\left(\lambda_{2}\right)+\frac{1}{3}\int_{\bar{\rho}(\lambda_{2})}^{\rho(\lambda_{2},x^{i})}\frac{d\rho}{\rho+p(\rho)}\,,\label{eq:conservation-1}
\end{equation}
then, the quantity
\begin{equation}
\tilde{\zeta}=-\frac{1}{6}\left(N+4\nu\right)-\frac{1}{3}\int_{\bar{\rho}(\lambda)}^{\rho(\lambda,x^{i})}\frac{d\rho}{\rho+p(\rho)}\,,\label{eq:conservation1}
\end{equation}
is conserved. One interesting aspect of Eq. \eqref{eq:conservation1}
is that the geometrical terms $N$ and $\nu$, present on the conservation
of $\tilde{\zeta}$, are precisely the same terms that contribute to
the linearized angular distance-redshift relation for the linearized
GLC gauge \cite{Fanizza:2020xtv}.

Let us now link our results to the standard perturbation theory. We follow the notation of \cite{Fanizza:2020xtv,Fanizza:2023ixk}, where the standard metric is given by
\begin{align}
ds^{2}&=\,a^{2}\left[-\left(1+2\phi\right)d\eta^{2}-2\mathcal{B}_{i}dx^{i}d\eta+\left(\bar{\gamma}_{ij}+\mathcal{C}_{ij}\right)dx^{i}dx^{j}\right]\,,\label{eq:std-metric}
\end{align}
and the SVT decomposition given by
\begin{align}
\mathcal{B}_{i}&=B_{i}+\partial_{i}B\,,\nonumber\\
C_{ij}&=\,-2\bar{\gamma}_{ij}\psi+2D_{ij}E+2\nabla_{(i}F_{j)}+2h_{ij}\,,\label{eq:std-pt}
\end{align}
where $B_{i}$ and $F_{i}$ are divergenceless vectors and $h_{ij}$ is a trace-less and divergenceless tensor. We also have
\begin{align}
    D_{ij}E=\left(\nabla_{(i}\nabla_{j)}-\bar{\gamma}_{ij}\frac{\Delta_{3}}{3}\right)E\,.\label{eq:decompDij}
\end{align}
In this case, the trace of $g_{ij}$ is proportional to $-\psi$, and then by using the relation between standard and GLC perturbations given by \cite{Fanizza:2020xtv,Fanizza:2023ixk}
\begin{equation}
\psi=-\frac{1}{6}\left(N+4\nu\right)\,,\label{eq:psi-GLC}
\end{equation}
we get that
\begin{equation}
\tilde{\zeta}=\psi-\frac{1}{3}\int_{\bar{\rho}(\lambda)}^{\rho(\lambda,x^{i})}\frac{d\rho}{\rho+p(\rho)}\,,\label{eq:conservation3}
\end{equation}
or, equivalently, with the use of Eq.~\eqref{eq:expansion},
\begin{equation}
\tilde{\zeta}
=\psi-\frac{1}{3}\int_{\bar{\rho}(\lambda)}^{\rho(1+\delta\rho)}\frac{d\rho}{\rho+p(\rho)}
\approx\psi-\frac{1}{3}\frac{\bar{\rho}\delta\rho}{\rho+p(\rho)}\,,\label{eq:conservation2}
\end{equation}
where we have expanded at linear order in the density perturbations in the last equality. Since we are working at first order in the gradient expansion, where the spatial gauge modes occur at the next-to-leading order, and the time gauge mode is fixed,
the quantity $\tilde{\zeta}$ given in Eq.~\eqref{eq:conservation1} is gauge invariant. Hence, within this approximation scheme, we may identify it with the curvature perturbation $\zeta$. For the complete expression of $\zeta$ to order $\mathcal{O}(\delta,\epsilon^{n})$, see Eqs. (2.27) of \cite{Fanizza:2023ixk}, where we also provide its gauge invariance proof in terms of light-cone perturbations.

\subsection{Non-linear $\zeta$}

As we have seen, our non-linear SU approach in the GLC gauge allowed us to obtain Eq. \eqref{eq:conservation_gradient} by neglecting the last term in Eq.~\eqref{eq:conservation}, which we have shown to correspond to the terms neglected in the standard perturbation theory \cite{Wands:2000dp} at linear order. Now, with the aim of going beyond this result, we leave the lapse function $\mathcal{M}$ unspecified and use the SU approach on the light-cone to prove the non-linear conservation of the curvature perturbation in terms of LC parameters. Just as done in the previous subsection, we start from Eq.~\eqref{eq:conserv} which gives
\begin{align}
\Theta_{u}=-\frac{1}{\left(\rho+p\right)}\frac{d\rho}{d\lambda}\,.
\label{eq:energyconserv-2}
\end{align}
Moreover, we consider Eq.~\eqref{eq:K-hij-trace} and the approximation exploited after Eq.~\eqref{theta-GLC1}, namely $\Theta_{n}\equiv\nabla_{\mu}n^{\mu}\simeq \Theta_{u}$. We then get that
\begin{align}
\frac{1}{\Upsilon\sqrt{\gamma}}\frac{d\left(\Upsilon\sqrt{\gamma}\right)}{d\lambda}=-\frac{1}{\left(\rho+p\right)}\frac{d\rho}{d\lambda}\,.
\label{eq:non-linear-1}
\end{align}
We now integrate this equation along $\lambda$ and consider that the pressure is adiabatic. In this way, one obtains that
\begin{align}
\ln\left[\frac{\left(\Upsilon\sqrt{\gamma}\right)_{\lambda_{2}}}{\left(\Upsilon\sqrt{\gamma}\right)_{\lambda_{1}}}\right]=-\int_{\rho\left(\lambda_{1},x^i\right)}^{\rho\left(\lambda_{2},x^i\right)}\frac{d\rho}{\rho+p\left(\rho\right)}\,.
\label{eq:conserv1}
\end{align}

At this point, the r.h.s. can be manipulated in the same spirit of what done in Eq.~\eqref{eq:procedure} (see also \cite{Wands:2000dp}). By doing so, we get that
\begin{align}
\int_{\rho(\lambda_{1},x^{i})}^{\rho(\lambda_{2},x^{i})}\frac{d\rho}{\rho+p(\rho)}&=\int_{\bar{\rho}(\lambda_{2})}^{\rho(\lambda_{2},x^{i})}\frac{d\rho}{\rho+p(\rho)}-\int_{\bar{\rho}(\lambda_{1})}^{\rho(\lambda_{1},x^{i})}\frac{d\rho}{\rho+p(\rho)}+\int_{\bar{\rho}(\lambda_{1})}^{\bar{\rho}(\lambda_{2})}\frac{d\rho}{\rho+p(\rho)}\nonumber\\
&=\int_{\bar{\rho}(\lambda_{2})}^{\rho(\lambda_{2},x^{i})}\frac{d\rho}{\rho+p(\rho)}-\int_{\bar{\rho}(\lambda_{1})}^{\rho(\lambda_{1},x^{i})}\frac{d\rho}{\rho+p(\rho)}-\ln\left(\frac{\bar{\Upsilon}\sqrt{\bar{\gamma}}|_{\lambda_{2}}}{\bar{\Upsilon}\sqrt{\bar{\gamma}}|_{\lambda_{1}}}\right)\,,
\label{eq:procedure2}
\end{align}
where, from the first to the second line, we have used Eq.~\eqref{eq:conserv1} at the background level. Now, thanks to Eq.~\eqref{eq:procedure2}, we can rewrite Eq.~\eqref{eq:conserv1} as
\begin{align}
\ln\left(\frac{\Upsilon\sqrt{{\gamma}}}{\bar{\Upsilon}\sqrt{\bar{\gamma}}}\right)_{\lambda_{2}}+\int_{\bar{\rho}(\lambda_{2})}^{\rho(\lambda_{2},x^{i})}\frac{d\rho}{\rho+p(\rho)}=\ln\left(\frac{\Upsilon\sqrt{\gamma}}{\bar{\Upsilon}\sqrt{\bar{\gamma}}}\right)_{\lambda_{1}}+\int_{\bar{\rho}(\lambda_{1})}^{\rho(\lambda_{1},x^{i})}\frac{d\rho}{\rho+p(\rho)}\,.
\label{eq:conservation-non-linear}
\end{align}
This shows that there is a conserved quantity at first order in the gradient expansion. This quantity corresponds to the non-linear curvature perturbation $\zeta$
\begin{align}
\zeta=\frac{1}{3}\ln\left(\frac{\Upsilon\sqrt{{\gamma}}}{\bar{\Upsilon}\sqrt{\bar{\gamma}}}\right)+\frac{1}{3}\int_{\bar{\rho}(t)}^{\rho(t,x^{i})}\frac{d\rho}{\rho+p(\rho)}+\mathcal{O}\left(\epsilon^{2}\right)\,.\label{eq:non-linear-zeta}
\end{align}
which then generalizes the linear result in Eq.~\eqref{eq:conservation1}. 

Thus, the curvature perturbation defined in Eq.~\eqref{eq:non-linear-zeta} generalizes at the non-linear level, but at the first order in the gradient expansion, the expression of the gauge invariant curvature perturbation in the LC gauge formalism given in \cite{Fanizza:2023ixk}. In fact, the first term corresponds to the curvature perturbations whilst the second term corresponds to density perturbations in the same spirit as done in Eq. \eqref{eq:conservation3}. Finally, we remark that we have obtained Eq. \eqref{eq:non-linear-zeta} without specifying the lapse function $\mathcal{M}$, therefore we
still have the freedom to fix the time gauge mode, as we will see better later.

\subsection{$\delta\mathcal{N}$ formalism on the LC gauge}

Let us begin by writing explicitly the exact expression for the expansion rate defined by the normal vector $n^{\mu}$ given by Eq.~\eqref{eq:orthog}. This will then be applied to the evaluation of the number of e-folds using the SU picture of the LC gauge. Such approach will allow us to obtain a generalization of the non-linear $\delta\mathcal{N}$ formalism in terms of LC gauge metric entries. The expansion rate in the LC gauge is given by
\begin{equation}
\Theta_n=\nabla_{\mu}n^{\mu}=\frac{1}{\sqrt{-g}}\partial_{\mu}\left(n^{\mu}\sqrt{-g}\right)=\frac{1}{\mathcal{M}\Upsilon\sqrt{\gamma}}\frac{d}{d\lambda}\left(\mathcal{M}\Upsilon\sqrt{\gamma}\right)+\partial_{\mu}n^{\mu}\,,\label{eq:non-linear-LC-exp}
\end{equation}
where, from Eqs.~\eqref{eq:orthog} and \eqref{eq:shift-v-GLC-cond}, we have that
\begin{equation}
\partial_{\mu}n^{\mu}=-\frac{\partial_{t}\mathcal{M}}{\mathcal{M}^{2}}-\partial_{i}\left(\frac{N^{i}_{LC}}{\mathcal{M}}\right)=-\frac{\partial_{t}\mathcal{M}}{\mathcal{M}^{2}}+\mathcal{O}\left(\epsilon^{2}\right)\,.\label{eq:div-n}
\end{equation}
Now, using Eqs.~\eqref{eq:non-linear-LC-exp} and \eqref{eq:div-n}, we obtain
\begin{equation}
\Theta_n
=\frac{1}{\Upsilon\sqrt{\gamma}}\frac{d}{d\lambda}\left(\Upsilon\sqrt{\gamma}\right)+\mathcal{O}\left(\epsilon^{2}\right)\,.\label{eq:exp-rate-gradexp}
\end{equation}
An interesting thing about this result is that it is invariant in form
on the past light-cone, i.e. all the dependence on the lapse function
is hidden in $n^{\mu}\partial_{\mu}\equiv\frac{d}{d\lambda}$. Let us now integrate Eq. \eqref{eq:exp-rate-gradexp} to compute the non-linear
number of e-folds at first order in the gradient expansion in terms of light-cone entries. We can then easily obtain the following result
\begin{equation}
\mathcal{N}\left(\lambda_{f},\lambda_{i}\right)\equiv\frac{1}{3}\int_{\lambda_{i}}^{\lambda_{f}}\Theta_n d\lambda'=\frac{1}{3} \ln\left[\frac{\left(\Upsilon\sqrt{\gamma}\right)_{\lambda_f}}{\left(\Upsilon\sqrt{\gamma}\right)_{\lambda_i}}\right]+\mathcal{O}\left(\epsilon^{2}\right)\,.\label{eq:non-linear-e-folds}
\end{equation}

Let us note that $\mathcal{N}(\lambda_{f},\,\lambda_{i})$ from Eq.~\eqref{eq:non-linear-e-folds} is a biscalar, i.e. depends on the gauge fixing both at the initial and final slicing. One possible fixing of the lapse function is given by the uniform curvature light-cone (UCLC) gauge: in this case the effective local scale factor is given by its background value\footnote{We will be using the subscript UC to describe the UCLC gauge fixing and the subscript UD to describe the UDLC gauge fixing.}
\begin{align}
\left(\Upsilon\sqrt{\gamma}\right)_{UC}=\bar{\Upsilon}\sqrt{\bar{\gamma}}\,.\label{eq:effectivelocalUCLC}
\end{align} 
Therefore, if we fix both initial and final slices on the UCLC gauge, the number of e-folds will be given by its background value, as
\begin{equation}
\mathcal{N}_{UC}\left(\lambda_{f},\,\lambda_{i}\right)
=\bar{\mathcal{N}}\left(\lambda_{f},\,\lambda_{i}\right)\,.\label{eq:e-foldingn}
\end{equation}

Let us also introduce the uniform density light-cone (UDLC) gauge defined by
\begin{align}
\rho_{UD}\left(\lambda,\,x^{i}\right)=\bar{\rho}\left(\lambda\right)\,.\label{eq:UD-gauge-cond}
\end{align}
Then, within the UDLC gauge, we have
\begin{equation}
    \zeta=\,\frac{1}{3}\ln\left[\frac{\left(\Upsilon\sqrt{\gamma}\right)_{\lambda_{f\,UD}}}{\left(\bar{\Upsilon}\sqrt{\bar{\gamma}}\right)_{\lambda_{i}}}\right]
    =\,\frac{1}{3}\ln\left[\frac{\left(\Upsilon\sqrt{\gamma}\right)_{\lambda_{f\,UD}}}{\left(\Upsilon\sqrt{\gamma}\right)_{\lambda_{i\,UC}}}\right]\,\label{eq:zeta-udlc}
\end{equation}
where from the first to the second equality we have used Eq. \eqref{eq:effectivelocalUCLC}. We can now compare Eqs.~\eqref{eq:non-linear-e-folds} and~\eqref{eq:zeta-udlc}, and obtain that $\zeta$ can be related to the number of e-folds in the following way
\begin{align}
\mathcal{N}\left(\lambda_{f\,UD},\,\lambda_{i\,UC}\right)
=&\frac{1}{3}\ln\left[\frac{\left(\Upsilon\sqrt{\gamma}\right)_{\lambda_{f\,UD}}}{\left(\Upsilon\sqrt{\gamma}\right)_{\lambda_{i\,UC}}}\right]\nonumber\\
=&\frac{1}{3}\ln\left[\frac{\left(\Upsilon\sqrt{\gamma}\right)_{\lambda_{f\,UD}}}{\left(\bar{\Upsilon}\sqrt{\bar{\gamma}}\right)_{\lambda_f}}\frac{\left(\bar{\Upsilon}\sqrt{\bar{\gamma}}\right)_{\lambda_i}}{\left(\Upsilon\sqrt{\gamma}\right)_{\lambda_{i\,UC}}}\right]+\bar{\mathcal{N}}\left(\lambda_{f},\lambda_{i}\right)\nonumber\\
=&\frac{1}{3}\ln\left(\frac{\Upsilon\sqrt{\gamma}}{\bar{\Upsilon}\sqrt{\bar{\gamma}}}\right)_{\lambda_{f\,UD}}
+\bar{\mathcal{N}}\left(\lambda_{f},\lambda_{i}\right)\nonumber\\
=&-\zeta(\lambda_{f})+\bar{\mathcal{N}}\left(\lambda_{f},\lambda_{i}\right)\,.\label{eq:e-foldingnUDLC-UCLC} 
\end{align}
Hence, if we define
\begin{equation}  
\delta\mathcal{N}\equiv\,\mathcal{N}\left(\lambda_{f\,UD},\,\lambda_{i\,UC}\right)-\mathcal{N}\left(\lambda_{f\,UC},\,\lambda_{i\,UC}\right)\,,
\label{eq:deltaN-lightcone}
\end{equation}
we straightforwardly get, from ~\eqref{eq:e-foldingn} and~\eqref{eq:e-foldingnUDLC-UCLC}, that $\delta\mathcal{N}=-\zeta$, which is valid at any order in perturbation theory, in agreement with \cite{Sugiyama:2012tj}.
As a remark, we underline that, while Eq.~\eqref{eq:non-linear-e-folds} depends directly on the initial
and final slice, Eq.~\eqref{eq:deltaN-lightcone} depends only on the difference of the perturbations of the e-fold number in the UDLC and UCLC on the final slices. Now we will give an example of how this last result can be used to evaluate the power spectrum of $\zeta$ in terms of the LC metric entries. Following the procedure used in \cite{Sugiyama:2012tj}, we fix the UCLC gauge and adopt the SU approximation. In this way, the density perturbations can be written in terms of the background density with perturbed initial conditions as follows
\begin{align}
    \rho_{UC}(\mathcal{N}_{UC},\, \mathbf{x})=\bar{\rho}(\mathcal{\bar{N}},\, \varphi_{*}^{A}(x))\,,\label{eq:rho-UCLC}
\end{align} 
where we use Eq.~\eqref{eq:e-foldingn}. Also, $\varphi_{*}^{A}$ is the field content of the underlying inflationary model evaluated just after the horizon exit. The index $A=1,...,d$ refers to the possibility that inflation could happen with $d$ scalar fields. Instead, by considering the UDLC gauge, we would have
\begin{align}
\rho_{UD}(\mathcal{N}_{UD},\, \mathbf{x})=\bar{\rho}(\mathcal{N}_{UD})\,.\label{eq:rho-UDLC}
\end{align}
Since $\rho$ is a scalar, we can then write 
\begin{align}
\rho'(\mathcal{N}',\,x')=\rho(\mathcal{N},\,x)\,,\label{eq:scalarcondition}
\end{align} 
which holds between two generic sets of coordinates $\mathcal{N},\,x$ and $\mathcal{N}',\,x'$ i.e. the value of a scalar function in a given physical point does not depend on the choice of coordinates.

Using Eqs. \eqref{eq:rho-UCLC} and \eqref{eq:rho-UDLC} on Eq.~\eqref{eq:scalarcondition}, we get
\begin{align}
\bar{\rho}(\bar{\mathcal{N}},\, \varphi_{*}^{A}(x))= \bar{\rho}(\mathcal{N}_{UD})\,,\label{eq:toinvert}
\end{align}
where the $x'=x$ from Eq.~\eqref{eq:scalarcondition} corresponds to the choice of the spatial threading to fix the LC gauge. Also, we choose $\mathcal{N}'=\mathcal{N}_{UC}$, and $\mathcal{N}=\mathcal{N}_{UD}$. Then,  Eq.~\eqref{eq:toinvert} can be inverted as
\begin{align}
\mathcal{N}_{UD}=\bar{\mathcal{N}}(\bar{\rho},\,\varphi_{*}^{A}(\mathbf{x}))\,.\label{eq:toexpand}
\end{align}
Hence, by expanding the fields in Eq.~\eqref{eq:toexpand} at linear order as $\varphi^A_*=\bar\varphi^A_*+\delta\varphi^A_*$, using Eq. \eqref{eq:deltaN-lightcone}, we obtain
\begin{align}
\frac{1}{3}\ln\left[\frac{\left(\Upsilon\sqrt{\gamma}\right)_{UD}}{\left(\Upsilon\sqrt{\gamma}\right)_{UC}}\right]= & \mathcal{N}(\bar{\rho},\, \varphi_{*}^{A}(\mathbf{x}))|_{UD} - \bar{\mathcal{N}}(\bar{\rho})|_{UC}\nonumber\\
= & \delta\varphi_{*}^{A}\partial_{A}\bar{\mathcal{N}}+\frac{1}{2}\delta\varphi_{*}^{A}\delta\varphi_{*}^{B}\partial_{A}\partial_{B}\bar{\mathcal{N}}+...\,.\label{eq:deltaN-expansion}
\end{align}
Therefore, given an inflationary model one may link the value of $\varphi_{*}^{A}$ to the left hand side of Eq. \eqref{eq:deltaN-expansion}.

Altogether, Eq. \eqref{eq:deltaN-expansion} can be a starting point for the evaluation of $f_{NL}$ in terms of light-cone perturbations (see, for example, \cite{Lyth:2005fi} for an evaluation of $f_{NL}$ using the $\delta\mathcal{N}$ formalism). 
Therefore, the results presented in Eqs. \eqref{eq:deltaN-lightcone} and \eqref{eq:deltaN-expansion}
constitute a further step to obtain non-Gaussian predictions on the past light-cone, from the primordial universe and directly in terms of the metric entries.


\section{Conclusions}
\label{Conclusions}

In this manuscript we have developed a separate universe (SU) description in the non-linear LC gauge. We provide the non-linear conditions to fix the LC and the GLC gauges in terms of standard coordinates on the ADM formalism. The main difference with the previous works \cite{Lyth:2004gb,Sugiyama:2012tj,Talebian-Ashkezari:2016llx},
where the SU was considered,
is that for the LC and GLC gauges we cannot neglect the shift vector, since this contains information about inhomogeneities along the world-line.

As an application of our results, and a consistency check, we repeated the procedures of \cite{Wands:2000dp} and \cite{Lyth:2004gb} to prove the super-horizon conservation of the comoving curvature perturbation $\zeta$, when a light-like foliation of spacetime is taken. This conservation has been achieved by neglecting the non-adiabatic pressure within the SU scheme.

We then generalize the $\delta\mathcal{N}$ formalism \cite{Salopek:1990jq,Sasaki:1995aw,Sasaki:1998ug,Lyth:2004gb,Sugiyama:2012tj,Starobinsky:1982ee,Starobinsky:1986fxa,Lyth:2005fi}, in terms of the combination of the LC metric entries $\Upsilon\sqrt{\gamma}$ within the uniform density light-cone gauge which is one of our most important results.
Let us remark that the gradient expansion employed here simplifies the expression of the expansion rate at the non-linear level (see Eq.~\eqref{eq:non-linear-LC-exp}). This could help in simplifying also the perturbative expressions (see, for instance, the one presented in Eq.~(6.11) of \cite{Frob:2021ore}). 

The separate universe formalism provides a procedure to investigate the evolution of the perturbations for different inflationary models. The extension of the $\delta\mathcal{N}$ formalism on the past light-cone, as developed in this manuscript, allows the evaluation of such dynamical evolution directly over the past light-cone. This moves us one step forward to the evaluation of non-linear effects (such as backreaction effects and non-Gaussianities) since the primordial universe until the late-time one along such past light-cone. Indeed, as a future step, we aim to investigate primordial backreaction effects on different expansion rates using the above-mentioned formalism and well-posed averaging procedures on the past light-cone \cite{Gasperini:2011us,Fanizza:2019pfp}.

Finally, for what regards the possible non-Gaussianities associated to any inflationary model, the $\delta\mathcal{N}$ formalism is a very useful tool. In fact, as shown in \cite{Lyth:2005fi}, this formalism provides very simple expressions for $f_{NL}$ in terms of $\mathcal{N}$. 
In other words, the overall goal of the research program is to obtain a formalism to compute the curvature perturbations at horizon re-entry expressed in terms of light-cone entries. We remark this point since the subsequent evolution of this metric entries could then be compared to late-time expression of cosmological observables as, for instance, the ones presented in Eq.~\eqref{eq:observables}. This would provide a self-consistent framework entirely given on the light-cone to disentangle the primordial non-Gaussianities from the ones naturally emerging during the non-linear late-time dynamics \cite{Kehagias:2015tda,DiDio:2016gpd,Schiavone:2023olz}.


\section*{Acknowledgement}
We are thankful to Danilo Artigas for useful comments on the manuscript. GM and MM are supported in part by INFN under the program TAsP ({\it Theoretical Astroparticle Physics}). This work was partially supported by the research grant number 2022E2J4RK ``PANTHEON: Perspectives in Astroparticle and
Neutrino THEory with Old and New messengers'' under the program PRIN 2022 funded by the Italian Ministero dell?Universit\`a e della Ricerca (MUR) and by the European Union - Next Generation EU. GF and MM are supported by the FCT through the research project with ref. number PTDC/FIS-AST/0054/2021. GF is also member of the Gruppo Nazionale per la Fisica Matematica (GNFM) of the Istituto Nazionale di Alta Matematica (INdAM).

\begin{appendices}

\section{Linear $\delta\Ncal$ formalism on the light-cone}
\label{secNvsGLC}
In this appendix we derive the $\delta\mathcal{N}$ formalism at linear order in perturbation theory using the framework developed in \cite{Fanizza:2020xtv, Fanizza:2023ixk} and reviewed in Sect.~\ref{deltaN}. In this way we explicitly show the consistency of our results. 

To begin, we linearize Eq.~\eqref{eq:deltaN-lightcone} and obtain
\begin{align} 
\delta\mathcal{N}= & \frac{1}{3}\ln\left[\frac{\left(\Upsilon\sqrt{\gamma}\right)_{UD}}{\left(\Upsilon\sqrt{\gamma}\right)_{UC}}\right]
+\mathcal{O}(\epsilon^{2})\nonumber\\
= & \frac{1}{3}\ln\left[\frac{\left(\bar{\Upsilon}\sqrt{\bar{\gamma}}\right)(1+\delta\Upsilon)(1+2\nu)_{UD}}{\left(\Upsilon\sqrt{\gamma}\right)_{UC}}\right]
+\mathcal{O}(\delta^{2},\epsilon^{2})\nonumber\\
= & \frac{1}{3}\ln\left[1+\left(\delta\Upsilon+2\nu\right)_{UD}\right]+\mathcal{O}(\delta^{2},\epsilon^{2})\nonumber\,\\
= &\frac{1}{3}\left(\delta\Upsilon+2\nu\right)_{UD}+\mathcal{O}(\delta^{2},\epsilon^{2})\label{eq:e-foldnumber}
\end{align}
where we have defined 
\begin{equation}
    \Upsilon=\bar{\Upsilon}(1+\delta\Upsilon)
\end{equation}
and we recall that $(\Upsilon\sqrt{\gamma})_{UC}$ is equal to the background value, being $\psi=0$ within the uniform curvature gauge. Also, we have used the metric in Eq.~\eqref{eq:gen_pert_GLC_metric} and the scalar/pseudoscalar decomposition of Eq.~\eqref{eq:scalar-pseudo-scalar}. Since $\delta\Upsilon=N/2$, we then have that 
\begin{equation}
\delta\Ncal(\lambda_{1,}\lambda_{2,}x^{i})=\frac{1}{6}\left(N+4\nu\right)_{UD}
+\mathcal{O}(\delta^{2},\epsilon^{2})\,.\label{eq:GLC_deltaN}
\end{equation}
From the relation between the light-cone perturbation and the standard ones in Eq.~\eqref{eq:psi-GLC}, (see also  \cite{Fanizza:2023ixk}), one gets that
\begin{align}
    \psi=-\frac{1}{6}(N+4\nu)\,.\label{eq:psi-LC}
\end{align}
Therefore, Eq.~\eqref{eq:GLC_deltaN} together with Eq.~\eqref{eq:psi-LC} and the fact that $\psi_{UD}=\zeta$, gives the well known relation $\delta\mathcal{N}=-\zeta$ \cite{Sugiyama:2012tj}. 

The result obtained in Eq.~\eqref{eq:GLC_deltaN} proves that the $\delta\mathcal{N}$ formalism on the past light-cone is consistent with the light-cone perturbation theory developed in \cite{Fanizza:2020xtv, Fanizza:2023ixk}. Thereby, the $\delta\mathcal{N}$ formalism within the past light-cone framework, at linear order in perturbation theory, could also be obtained by starting from the results presented in Sect. \ref{deltaN}. To this aim, one should integrate Eqs.~\eqref{theta-GLC1} and~\eqref{eq:linear-theta}, evaluating them between the uniform curvature and the uniform density slices. 

\end{appendices}

\bibliographystyle{sn-mathphys-num.bst}
\bibliography{ref_DeltaN.bib}

\end{document}